\documentclass{article}

\begin{document}

\begin{center}
\centerline{\large \bf Comment on "Entropy production and the arrow of time"} 
\end{center}

\vspace{3 pt}
\centerline{\sl V.A.Kuz'menko\footnote{Electronic 
address: kuzmenko@triniti.ru}}

\vspace{5 pt}

\centerline{\small \it Troitsk Institute for Innovation and Fusion 
Research,}
\centerline{\small \it Troitsk, Moscow region, 142190, Russian 
Federation.}

\vspace{5 pt}

\begin{abstract}

J.M.R. Parrondo at al. in arXive:0904.1573 continue numerous efforts to 
unify the concepts of the arrow of time and entropy production with the 
concept of time invariance in physics. This is a wrong way.
        
\vspace{5 pt}
\end{abstract}

\vspace{12 pt}

The discussed article [1] continues the numerous efforts of the theorists 
to explain a nature of the arrow of time and entropy production (the 
second low of thermodynamics) in the time-invariant physics. The base of 
this enormous work is the concept of time-reversibility of the low of 
physics. However, this is a great error. This concept is only an assumption. 
It does not have any reliable experimental proofs in quantum physics.

In contrast, the concept of inequality of forward and reversed processes 
has now some direct and great number of indirect experimental proofs [2, 3]. 
The differential cross-sections of forward and reversed transitions in 
optics can differ in many orders of magnitude. This is a real physical base 
of all nonlinear optics.

So, in this case the arrow of time and entropy production are quite natural 
properties of physics. The entropy is a quantitative material measure of a 
memory of quantum system about its initial state. And this memory is a 
result of inequality of forward and reversed processes in physics. 

A wide field here exists for a theorist's work. Unfortunately, they continue 
to work in the wrong direction.

\end{document}